\documentclass[aps,pra,showpacs,twocolumn,superscriptaddress,floatfix]{revtex4}
\usepackage{amsmath,amssymb,color}
\usepackage{lipsum}
\usepackage{grffile}
\usepackage{epsfig}
\usepackage{hyperref}
\usepackage[all]{hypcap}

\newcommand{\BH}{\text{BH}}

\usepackage[normalem]{ulem}

\definecolor{darkBlue}{rgb}{0,0.1,0.5}

\begin{document}

\newcommand{\lmu} {Department of Physics and Arnold Sommerfeld Center for Theoretical Physics,
Ludwig-Maximilians-Universit{\"a}t M{\"u}nchen, Theresienstr.\ 37, 80333 Munich, Germany}

\title{Domain-wall melting in ultracold boson systems with hole and spin-flip defects}
\author{Jad C. Halimeh} 
\affiliation{\lmu}
\author{Anton W\"ollert}
\affiliation{\lmu}
\author{Ian McCulloch}
\affiliation{Centre for Engineered Quantum Systems, School of Mathematics and Physics,
University of Queensland, Brisbane 4072, Australia}
\author{Ulrich Schollw\"ock} 
\affiliation{\lmu}
\author{Thomas Barthel} 
\affiliation{Laboratoire de Physique Th\'{e}orique et Mod\`{e}les Statistiques, Universit\'{e} Paris-Sud, CNRS UMR 8626, 91405 Orsay, France}
\affiliation{\lmu}

\date{June 6, 2013}

\begin{abstract}
Quantum magnetism is a fundamental phenomenon of nature. As of late, it has garnered a lot of interest because experiments with ultracold atomic gases in optical lattices could be used as a simulator for phenomena of magnetic systems. A paradigmatic example is the time evolution of a domain-wall state of a spin-$1/2$ Heisenberg chain, the so-called domain-wall melting. The model can be implemented by having two species of bosonic atoms with unity filling and strong on-site repulsion $U$ in an optical lattice. In this paper, we study the domain-wall melting in such a setup on the basis of the time-dependent density matrix renormalization group (tDMRG). We are particularly interested in the effects of defects that originate from an imperfect preparation of the initial state. Typical defects are holes (empty sites) and flipped spins. We show that the dominating effects of holes on observables like the spatially resolved magnetization can be taken account of by a linear combination of spatially shifted observables from the clean case. For sufficiently large $U$, further effects due to holes become negligible. In contrast, the effects of spin flips are more severe as their dynamics occur on the same time scale as that of the domain-wall melting itself. It is hence advisable to avoid preparation schemes that are based on spin-flips.
\end{abstract}


\maketitle

\section{Introduction}
Recent progress in ultracold-atomic-gas experiments \cite{Ketterle1999,Bloch2008} has allowed for greater degrees of control where now one has the tools to explore many interesting and fascinating phenomena of quantum many-body physics that previously have been restricted only to the realm of theoretical investigation. 
Gases of ultracold fermionic and bosonic atoms in optical lattices provide the arguably cleanest implementations of the Fermi- and Bose-Hubbard models and are very well tunable. These fundamental models of condensed matter physics have by now been studied quite extensively in diverse experiments. See for example Refs.\ \cite{Greiner2002,Modugno2002,Stoeferle2004,Jordens2008,Roati2008,Schneider2008,esslinger2010fermi,serwane2011deterministic,bloch2012quantum,schneider2012fermionic,Ronzheimer2013-110}.
The experimental capabilities are very well developed, as exemplified by the controlled shifting between the superfluid (SF) and Mott-insulator (MI) regimes \cite{Bloch2008,Greiner2002}, generation of random potentials \cite{SanchezPalencia2007,Billy2008,Kondov2011}, single-atom imaging \cite{Bakr2009,bakr2010probing,Sherson2010,Streed2012}, or single-site manipulation \cite{Wuert2009,Weitenberg2011}.

\begin{figure}[]
\includegraphics[width=\columnwidth]{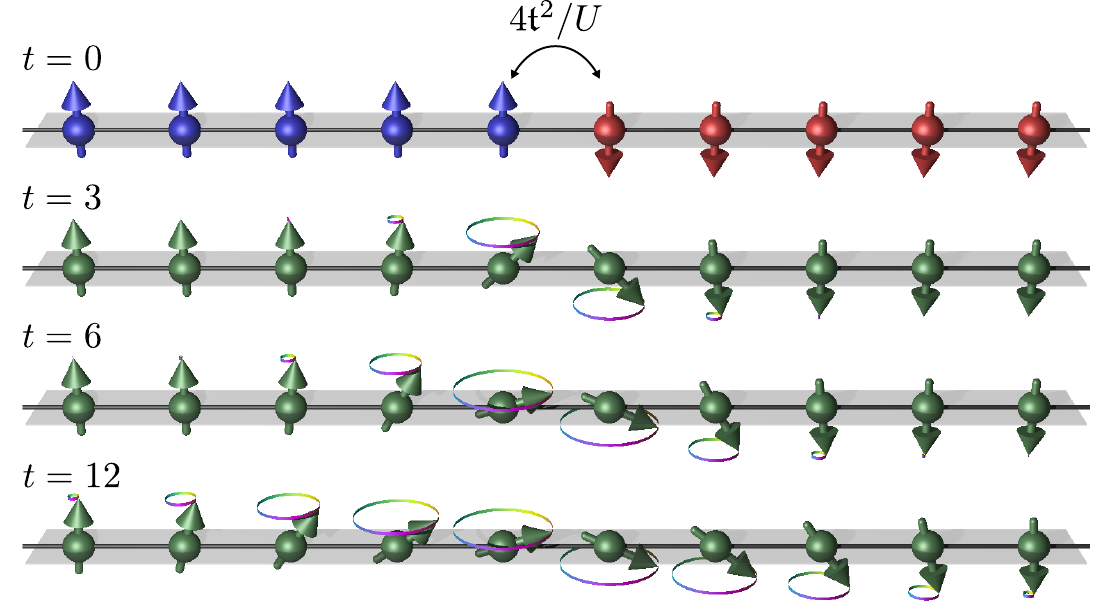}
\caption{(Color online)
Initial \emph{clean} domain-wall state with spin-up (spin-down) bosons on the left (right) half of the system at $t=0$. The illustrations for times $t>0$ are based on the actual evolution of the on-site magnetizations $\langle \hat S^z_j\rangle$ with hopping amplitude $\mathfrak{t}=1$ and onsite repulsion $U=15$. The domain wall melts and evolves into a nontrivial magnetization profile.
}
\label{fig:Fig1}
\end{figure}
\begin{figure*}[]
\includegraphics[width=0.9\textwidth]{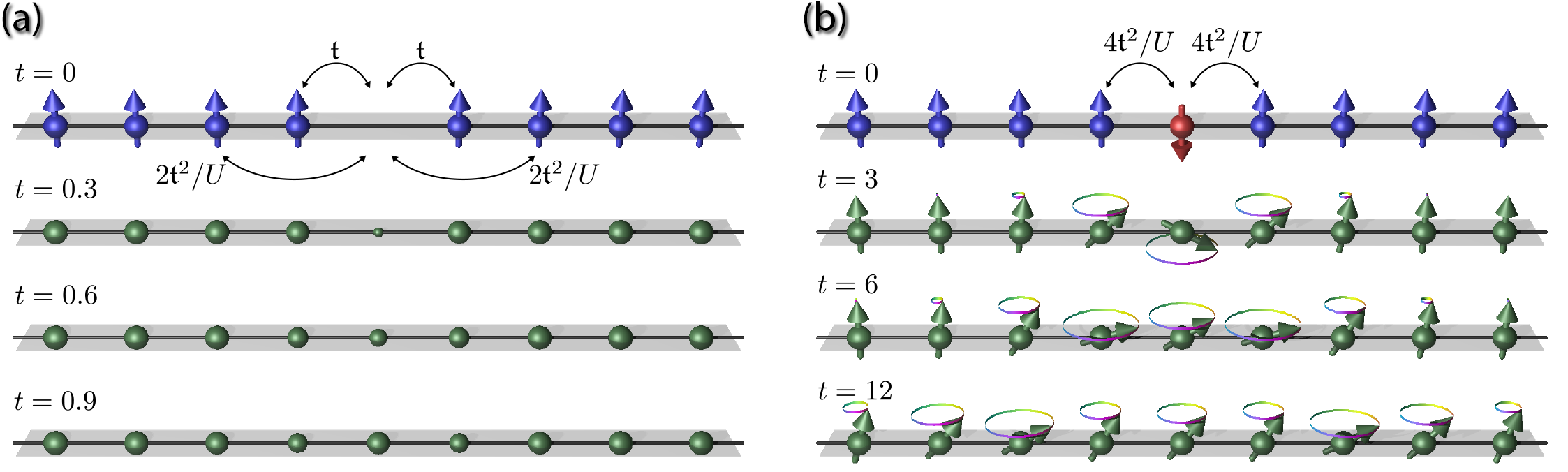}
\caption{(Color online)
Illustration of the time evolution of a hole and a spin-flip defect in a fully polarized background. The sketches are based on the actual evolution of densities $\langle \hat{n}_{\uparrow,j}+\hat{n}_{\downarrow,j}\rangle$ and magnetizations $\langle \hat S^z_j\rangle$ for hopping amplitude $\mathfrak{t}=1$ and onsite repulsion $U=15$. (a) Holes can move by one site via direct hopping and by two sites via a second-order process that gets suppressed with increasing $U$. As explained in the text, the density dynamics in the effective $\mathfrak{t}$-$J$ model \eqref{eq:tJ} is independent of the simultaneous spin dynamics. The latter is, however, influenced by density fluctuations. (b) The Heisenberg-type spin-spin interaction is caused by a second-order exchange process where bosons hop between nearest-neighbor sites. The focus of this paper is to explore the effects of such defects on the domain-wall melting in Fig.~\ref{fig:Fig1}.
}
\label{fig:Fig2}
\end{figure*}

In the vein of Feynman's idea to use one well-controllable quantum many-body system to simulate others \cite{Feynman1982,Buluta2009-326}, it is of particular interest to gain a thorough understanding of experimentally feasible ultracold-atomic-gas systems that can be used to faithfully implement spin models. Such setups could then be used to study the diverse phenomena of quantum magnetism.

As it turns out, the drosophila of quantum magnetism, the Heisenberg spin-$1/2$ XXZ model, appears quite naturally as an effective model for the subspace of unitary occupancy of the two-species Bose-Hubbard (BH) model in the limit of strong on-site interaction strengths \cite{Kuklov2003,Duan2003,Ripoll2003,Altman2003,Barthel2009}. The effective spin-exchange couplings are determined by the tunneling parameters and the inter- and intra-species interaction strengths.  Numerical investigations \cite{Barthel2009} have been presented and an experimental realization of this model has recently been implemented in order to study the quantum dynamics of a single spin-impurity \cite{Fukuhara2013}. One can envisage many interesting experiments using this setup in order to observe and investigate important many-body phenomena such as quantum phase transitions, long-range order, the temporal growth of entanglement, diffusive versus ballistic transport, relaxation dynamics, or integrability, to name a few. It can also provide a testbed for ultracold-atomic-gas experimental setups where their robustness to defects can be investigated and scrutinized.

A prominent nonequilibrium process that comprises several of the aforementioned many-body phenomena is the melting of a domain wall as depicted in Fig.~\ref{fig:Fig1}, and this naturally becomes an important phenomenon to probe in ultracold-atom experiments that aim to map onto the spin-$1/2$ XXZ model. 
Initially, the system is in a product state where the left half of the system is occupied by up-spins and the right half by down-spins. During the evolution, magnetization flows from left to right, accompanied by a growing entanglement. The dynamics has been studied analytically and numerically, for example, in Refs.\ \cite{Gochev1977-26,Gochev1983-58,Antal1999,Hunyadi2004,Gobert2005,Mossel2010-12,Jesenko2011-84,Zauner2012,Lancaster2010-81,Cai2011}. The transport is ballistic in the critical XY phase of the model. In the gapped phases, after some initial ballistic transport, the spin current was found to vanish for longer times. Besides this, there is an interesting nearest-neighbor beating effect in the magnetization profile (synchronized opposing oscillations of the magnetizations on neighboring sites) and small plateaus evolve at the domain-wall fronts. This can be attributed to the integrability of the model.
 
When one implements the domain-wall melting experimentally with ultracold bosons, defects can occur due to an imperfect preparation of the initial state. There are basically two options for the preparation. (i) In the first scheme, one prepares a Mott insulating state of spin-up bosons. Then, using a light mask, one addresses the right half of the system, bringing it into resonance with a microwave pulse that causes the spins to flip. (ii) In an alternative scheme, using light masks for the halves of the systems, one can cool the bosons in the lattice with strong chemical potential differences for the two species.
In both schemes, due to an ultimately finite temperature, hole defects can occur (Fig.~\ref{fig:Fig2}a). Due to the (shallow) trapping potential, these holes correspond to the lowest excitations of the Mott insulator ground state. The first scheme allows for the preparation of a spatially tighter (less smeared) domain wall \cite{Comm2013a}. One disadvantage of this scheme is the finite spin-flip efficiency (typically around 98\% in current experiments) which corresponds to the occurrence of spin-flip defects as shown in Fig.~\ref{fig:Fig2}b.

In this paper, numerical studies are presented for the domain-wall melting of two boson species in a one-dimensional optical lattice using various values of the on-site interaction strength $U$, most of which lie in the large-$U$ limit where the model maps faithfully onto a corresponding spin-$1/2$ XXZ model. All corresponding model parameters offer experimental feasibility and are simulated closely following the conditions in Fukuhara \emph{et al.} \cite{Fukuhara2013}, and are thus very relevant to similar future experimental investigations.
We focus in particular on the effects of typical experimental defects in the initial state on the melting dynamics. The quasi-exact numerical treatment is performed using the time-dependent density matrix renormalization group (tDMRG) method \cite{White1992,White1993,Vidal2004,White2004,Daley2004} in 
the Krylov approach \cite{Feiguin2005,Garcia-Ripoll2006-8,McCulloch2007-10} (see also \cite{Schmitteckert2004-70}). The results show that the dominating effects of hole defects on observables like the spatially resolved magnetization can be taken account of by a simple averaging procedure over spatially shifted observables from the clean case. To some extent, this smoothens out the beating and plateaus in the magnetization profile of the melting domain wall. For large $U$, the hole dynamics is much faster than the domain-wall dynamics. Hence, effects of holes beyond the aforementioned smoothening effect become negligible. In contrast, the effects of spin flips are more severe as their dynamics occur on the same time scale as that of the domain-wall melting itself. The spatial averaging procedure employed for the holes is still useful but not as powerful in this case. For the experimental investigations, this gives a reason to favor the second preparation scheme, cooling with chemical potentials, over the first scheme that is based on inducing spin-flips for one half of the system.

The paper is divided into five sections beyond the introduction: In Section II, the models occurring in this study and the mappings between them are discussed. After a specification of the different initial states in Section III, Section IV presents numerical simulations showing how the BH dynamics approaches the $\mathfrak{t}$-$J$ model dynamics. In Section V, the main results are presented and explained along with a discussion of the various observables of interest that are best suited to study the domain-wall melting. The paper concludes with Section VI and a convergence analysis of the numerical simulations in the appendix.

\section{Models}
\subsection{Spin-\texorpdfstring{$1/2$}{1/2} XXZ chain}\label{sec:XXZmodel}
The spin-$1/2$ XXZ Heisenberg magnet is a classic example of a one-dimensional quantum lattice model that has been extensively studied \cite{Giamarchi,Sutherland,Korepin} and that is of ideal importance to the understanding of magnetism and various phenomena in quantum many-body physics as mentioned in the introduction. Considering a one-dimensional lattice of $L$ sites, the Hamiltonian describing this model is
\begin{equation} \label{eq:XXZ}
\hat{H}_{\text{XXZ}}=J_{\perp}\sum_{j=1}^{L-1}(\hat{S}^x_j\hat{S}^x_{j+1}+\hat{S}^y_j\hat{S}^y_{j+1})
+J_z\sum_{j=1}^{L-1}\hat{S}^z_j\hat{S}^z_{j+1},
\end{equation}
where the spin operators obey the commutation relations $[\hat{S}_i^\alpha,\hat{S}_j^\beta]=i\delta_{ij}\epsilon_{\alpha\beta\gamma}\hat{S}_i^\gamma$ ($\hbar=1$).

The properties of the ground state of this Hamiltonian crucially depend upon the in-plane and on-axis spin-spin interaction parameters $J_{\perp}$ and $J_z$.  In the case $J_{\perp}=J_z$, $\hat{H}_{\text{XXZ}}$ becomes the isotropic Heisenberg Hamiltonian \cite{Heisenberg1928,Bethe1931} and the interaction between the spins is rotation-invariant.  When $J_{\perp},J_z>0$, the Hamiltonian is antiferromagnetic, since it is energetically favorable that the spins on neighboring sites have anti-parallel alignment, while when $J_{\perp},J_z<0$, parallel alignment is favorable and thus the Hamiltonian is ferromagnetic.  Moreover, at the critical point $J_z/|J_{\perp}|=1$, there is a Kosterlitz-Thouless-type phase transition that the system undergoes from a gapless XY regime ($J_z/|J_{\perp}|\leq1$) to the gapped ($J_z/|J_{\perp}|>1$) N\'{e}el phase.

Domain-wall melting in this system has been investigated analytically and numerically \cite{Gochev1977-26,Gochev1983-58,Antal1999,Hunyadi2004,Gobert2005,Mossel2010-12,Jesenko2011-84,Lancaster2010-81,Zauner2012}, and one can observe a transition from ballistic to subdiffusive dynamics when going from the gapless to the gapped regime.  To be able to simulate this model with an ultracold-atomic-gas system would be a very interesting way to experimentally probe such dynamics, and such a mapping has already been proposed \cite{Kuklov2003,Duan2003,Ripoll2003,Altman2003,Barthel2009}, where a two-species BH model in the limit of large interactions at unity filling can be approximated by the spin-$1/2$ XXZ model with an induced ordering field.  This is discussed in the following.

\subsection{Two-species Bose-Hubbard model and the relation to the XXZ model}
A prominent example for using bosonic systems to simulate others \cite{Feynman1982,Buluta2009-326} is that of using the \emph{two-species Bose-Hubbard} (BH) model (ultracold bosonic atoms in optical lattices) to emulate the spin-$1/2$ XXZ model \cite{Kuklov2003,Duan2003,Ripoll2003,Altman2003,Barthel2009}, where the two boson species correspond to spins up and down, respectively. This two-species BH model is described by the Hamiltonian
\begin{multline}\label{eq:BH}
\hat{H}_{\text{BH}} = -\sum_{\sigma,j=1}^{L-1}\mathfrak{t}_{\sigma}(\hat{b}^{\dagger}_{\sigma,j}\hat{b}_{\sigma,j+1} +h.c.)\\ 
+\sum_{\sigma,j=1}^{L}\frac{U_{\sigma}}{2}\hat{n}_{\sigma,j}(\hat{n}_{\sigma,j}-1)
+V\sum_{j=1}^{L}\hat{n}_{\uparrow,j}\hat{n}_{\downarrow,j},
\end{multline}
where $\sigma$ ($=\uparrow$ or $\downarrow$) labels the boson species, $\mathfrak{t}_{\sigma}$ is the tunneling parameter for `$\sigma$' bosons, $U_{\sigma}$ is the intra-species on-site interaction strength for `$\sigma$' bosons, $V$ is the inter-species interaction strength between `$\uparrow$' and `$\downarrow$' bosons on the same site, $\hat{b}_{\sigma,j}$ is the annihilation operator for `$\sigma$' bosons on site $j$ ($[\hat{b}_{\sigma,j},\hat{b}^\dag_{\sigma',j'}]=\delta_{\sigma\sigma'}\delta_{jj'}$), and $\hat{n}_{\sigma,j}=\hat{b}^{\dagger}_{\sigma,j}\hat{b}_{\sigma,j}$ is the number operator for `$\sigma$' bosons on site $j$. The bosonic species `$\uparrow$' and `$\downarrow$' are associated with two internal states of the atomic species used in the experimental setup (such as rubidium isotope $^{87}$Rb, where the two species correspond to two hyperfine states $|F=1,m_F=+1\rangle$ and $\left|F=2,m_F=-1\right\rangle$ of the bosonic atom).  Moreover, both bosonic species can be trapped by separate standing laser-light waves \emph{via} polarization selection \cite{Mandel2003} and the spin distribution of such a system can then be probed by single-site-resolved fluorescence imaging with a high-resolution microscope objective \cite{Bakr2009,bakr2010probing,Sherson2010}.

In the limit of large $U_\uparrow$, $U_\downarrow$, and $V$, using second-order perturbation theory or the corresponding Schrieffer-Wolff transformation \cite{Kuklov2003,Duan2003,Ripoll2003,Altman2003,Barthel2009}, one can derive an effective Hamiltonian for the subspace of unity filling (number of particles equal to the number of lattice sites), yielding
\begin{equation}
\hat{H}_{\text{XXZ}}-h\sum_{j=1}^{L}\hat{S}^z_j,
\end{equation}
where
\begin{align}
J_z &= 2\frac{\mathfrak{t}_{\uparrow}^2+\mathfrak{t}_{\downarrow}^2}{V}-\frac{4\mathfrak{t}_{\uparrow}^2}{U_{\uparrow}}-\frac{4\mathfrak{t}_{\downarrow}^2}{U_{\downarrow}},\quad
J_{\perp}=-\frac{4\mathfrak{t}_{\uparrow}\mathfrak{t}_{\downarrow}}{V}, \\
h &= \frac{4\mathfrak{t}_{\uparrow}^2}{U_{\uparrow}}-\frac{4\mathfrak{t}_{\downarrow}^2}{U_{\downarrow}}.
\end{align}
The induced homogeneous magnetic field $h$ can be ignored due to the conservation of the total magnetization in Eq.~\eqref{eq:XXZ}. The spin-exchange terms are due to a second-order process where bosons hop twice between neighboring sites and the energy in the intermediate states is increased due to the on-site repulsion $U_\sigma$, $V$.

For the experimentally most relevant situation $\mathfrak{t}_{\uparrow}=\mathfrak{t}_{\downarrow}\equiv\mathfrak{t}$, and $U_{\uparrow}=U_{\downarrow}=V\equiv U$, one arrives at the isotropic Heisenberg antiferromagnet with $J_{\perp}=J_z$.
This regime is at the focus of this paper since, on the one hand, the main purpose of the paper is to study the effect of holes and spin flips on domain-wall melting rather than the effect of anisotropies on it and, on the other hand, significant anisotropies are very hard to achieve experimentally \cite{Fukuhara2013,Comm2013a}.  For instance, the variance in $V$ is typically given by the parameter
\begin{equation}
\Delta V = \frac{U_{\uparrow}+U_{\downarrow}}{2}-V
\end{equation}
and $\Delta V$ can be set experimentally \cite{Fukuhara2013,Comm2013a} to a value in $[-0.1,0.1]\times U_{\uparrow}$. Note that the available range for the effective spin couplings can be extended substantially by employing optical superlattices as discussed and demonstrated for example in Refs.~\cite{Foelling2007,Trotzky2008,Barthel2009}.

\subsection{The \texorpdfstring{$\mathfrak{t}$-$J$}{t-J} model as an effective model for strong repulsion}
 \begin{figure}[]
\includegraphics[width=0.925\columnwidth]{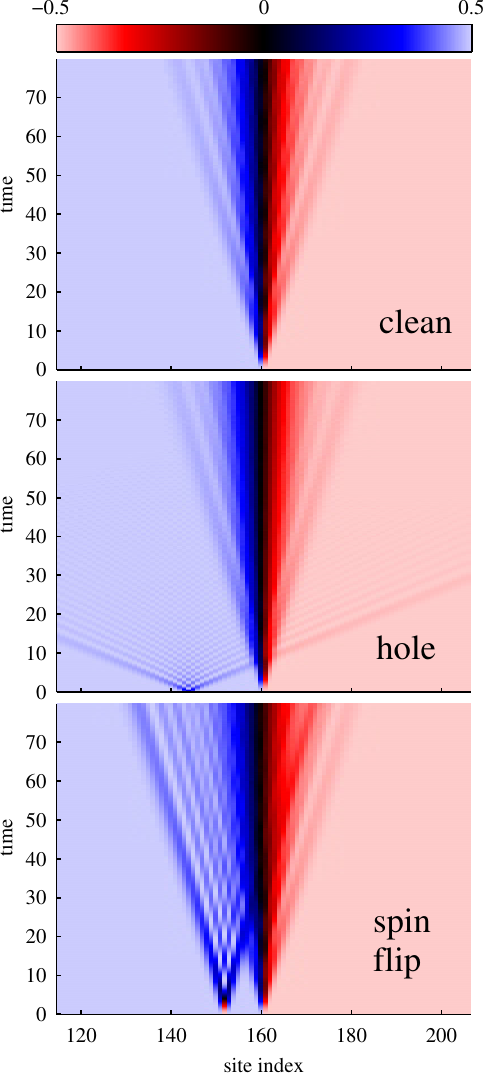}
\caption{(Color online)
The spin-density $\langle \hat{S}^z_j(t)\rangle$ as a function of position and time for $\mathfrak{t}=1$, $U=15$, lattice size $L=320$, and defect positions $j_h=L/2-16$, $j_f=L/2-8$. All three maps show times up to $80/\mathfrak{t}$ as by then the part of the hole that initially moves away from the domain wall will have been reflected off the boundary but still not interacted with the domain wall. Like the domain wall, the spin-flip evolves on a time scale $(4\mathfrak{t}^2/U)^{-1}$, while the hole defect moves on the shorter time scale $\mathfrak{t}^{-1}$. It is interesting to note in the cases of the clean domain-wall and spin-flip initial states that along the domain wall there is a nearest-neighbor \emph{beating} behavior (synchronized opposing oscillations on neighboring sites) that is absent in the hole case. See also Fig.~\ref{fig:Fig4} for slices and the Supplementary Material \cite{epaps} for animations.
}
\label{fig:Fig3}
\end{figure}
Since we want to study the effect of hole defects which occur in the experiments, we can not restrict the analysis to the subspace of unitary occupancy as done in the previous section. Rather, one has to take into account all states where on each site we have either one or no boson. The second-order perturbation theory for the limit of strong repulsion leads in this case to a bosonic variant of the so-called $\mathfrak{t}$-$J$ Hamiltonian \cite{Spalek1977,Chao1977,Chao1978,Anderson1987,Dagotto1994}, containing in this case some three-site terms that are particular to the bosonic nature of the particles. For our specific two-species Bose-Hubbard model \eqref{eq:BH}, we obtain a hard-core boson $\mathfrak{t}$-$J$ model
\begin{equation}\label{eq:tJ}
\hat{H}_{\mathfrak{t}\text{-}J}=\hat{H}_{\mathfrak{t}}+\hat{H}_{\text{XXZ}}+\hat{H}_{3-\text{site}},
\end{equation}
where $\hat{H}_{\text{XXZ}}$ is the XXZ Hamiltonian \eqref{eq:XXZ} that encodes the nearest-neighbor spin exchange and
\begin{equation}
\hat{H}_{\mathfrak{t}}=-\sum_{\sigma,j=1}^{L-1}\mathfrak{t}_{\sigma}(\hat{a}^{\dagger}_{\sigma,j}\hat{a}_{\sigma,j+1} +h.c.)
\end{equation}
is the direct hopping. Here, $\hat{a}_{\sigma,j}$ are hard-core-bosonic annihilation operators  with commutation relations $[\hat{a}_{\sigma,j},\hat{a}^\dag_{\sigma',j'}]=\delta_{\sigma\sigma'}\delta_{jj'}$ $\forall_{j\neq j'}$ and $\{\hat{a}_{\sigma,j},\hat{a}^\dag_{\sigma',j}\}=\delta_{\sigma\sigma'}$. In terms of the Pauli matrices $\{\hat{\sigma}^\alpha|\alpha=x,y,z\}$, the spin operators (occurring in $\hat{H}_{\text{XXZ}}$ and $\hat{H}_{3-\text{site}}$) are given by
\begin{equation}
\hat{S}^\alpha_j := \frac{1}{2}\sum_{\sigma\sigma'}\hat{a}^\dag_{\sigma,j}[\hat{\sigma}^\alpha]_{\sigma\sigma'}\hat{a}^{\phantom{\dag}}_{\sigma',j}.
\end{equation}
The terms
\begin{align} \nonumber
\hat{H}_{3-\text{site}}
=&-\sum_{\sigma,j=1}^{L-2}\frac{\mathfrak{t}_{\sigma}^2}{V}(\hat{a}^{\dagger}_{\sigma,j}\hat{n}_{-\sigma,j+1}\hat{a}_{\sigma,j+2}+h.c.) \\ \nonumber
&-\frac{\mathfrak{t}_{\uparrow}\mathfrak{t}_{\downarrow}}{V}\sum_{\sigma,j=1}^{L-2}(\hat{a}^{\dagger}_{-\sigma,j}\hat{S}^{\sigma}_{j+1}\hat{a}_{\sigma,j+2}+h.c.) \\
&-\sum_{\sigma,j=1}^{L-2}\frac{2\mathfrak{t}_{\sigma}^2}{U_{\sigma}}(\hat{a}^{\dagger}_{\sigma,j}\hat{n}_{\sigma,j+1}\hat{a}_{\sigma,j+2}+h.c.)
\end{align}
describe second-order processes, where bosons move by two sites. In the first term, a `$\sigma$' boson hops via a site occupied by a `$-\sigma$' boson. In the second term, a `$\sigma$' boson hops from site $j+2$ to a neighboring site $j+1$ occupied by a `$-\sigma$'. The latter subsequently hops to site $j$, causing an effective spin-flip on site $j+1$. In the third term, a `$\sigma$' boson hops over a site occupied by the same species to a next-nearest-neighbor site. We have used the notation $\hat{S}^{\sigma}_{j+1}$ to denote $\hat{S}^+_{j+1}$ ($\hat{S}^-_{j+1}$) when $\sigma$ is `$\uparrow$' (`$\downarrow$').

\section{Initial states}
\begin{figure}[t!]
\includegraphics[width=\columnwidth]{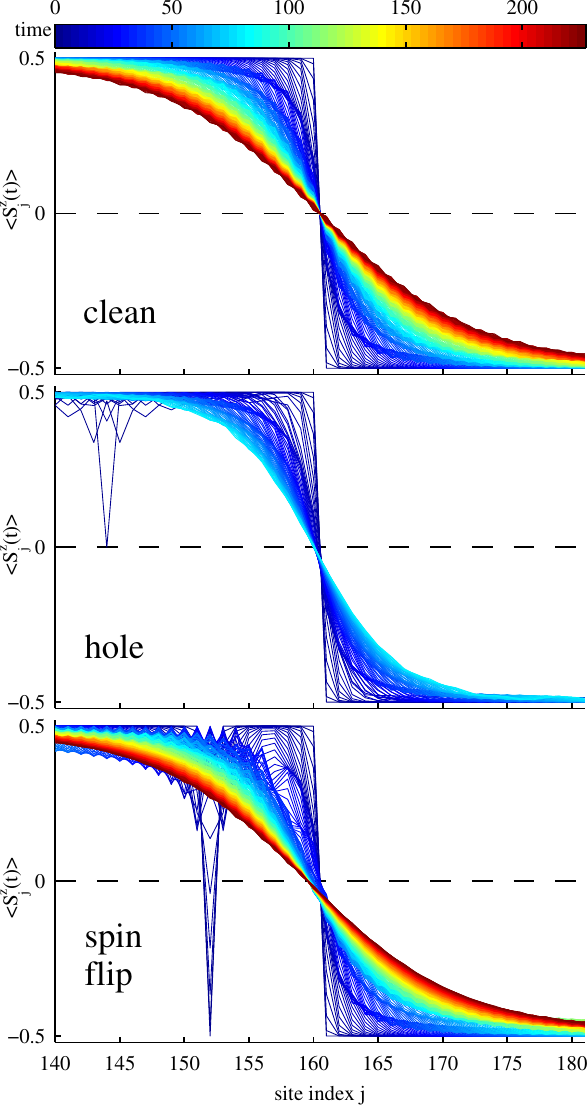}
\caption{(Color online)
Spin-density $\langle \hat{S}^z_j(t)\rangle$ profiles around the domain wall at various times for the clean, hole, and spin-flip initial states ($\mathfrak{t}=1$, $U=15$, $L=320$, $j_h=L/2-16$, $j_f=L/2-8$).  Time is indicated chromatically where blue corresponds to $t=0$ and red to $t=227/\mathfrak{t}$ (reached for the clean and spin-flip cases).  In the case of the hole, only times up to $t=80/\mathfrak{t}$ are shown. Also here one sees distinctive nearest-neighbor beating behavior in the clean and spin-flip cases that is considerably smoothened out in the hole case. Online, animations are provided that show the evolution of the states with defects in direct comparison to the clean domain-wall state \cite{epaps}.}
\label{fig:Fig4}
\end{figure}
In investigating the dynamics of a global quench where an initial state $|\psi_0\rangle=\left|\psi_0(t\leq 0)\right\rangle$ is time-evolved for $t>0$ with the Hamiltonian $\hat{H}$, which can be either $\hat{H}_{\text{BH}}$ or $\hat{H}_{\mathfrak{t}\text{-}J}$ for the purposes of this paper, it is particularly interesting to study the effect of \emph{defects} in the initial domain-wall state on the melting dynamics, because defects such as holes and spin flips can occur naturally in the preparation process. For our numerical investigations of the full BH model, the \emph{clean} domain-wall initial state $|\psi^\BH_c\rangle$ is chosen to be the ground state of the Hamiltonian
\begin{subequations}
\begin{gather}
\hat{H}_{\text{prep}}:=\hat{H}_{\text{BH}} -\mu\sum^{L/2}_{j=1}(\hat{n}_{\uparrow,j}+\hat{n}_{\downarrow,j+L/2})\\
\hat{H}_{\text{prep}}|\psi^\BH_c\rangle=E_0|\psi^\BH_c\rangle
\end{gather}
\end{subequations}
at unity filling with $L/2$ `$\uparrow$' bosons and $L/2$ `$\downarrow$' bosons on an $L$-site lattice. The species- and site-dependent chemical potential ($\mu$), when chosen sufficiently large compared to the hopping amplitude $\mathfrak{t}_\sigma$ in $\hat{H}_{\text{BH}}$, ensures that a domain-wall state is formed whereby the left half of the lattice ($1\leq j\leq L/2$) is mostly occupied by `$\uparrow$' bosons and the other half ($L/2+1\leq j\leq L$) is mostly occupied by `$\downarrow$' bosons. For large chemical potential and density-density interaction ($U_\sigma$, $V$), the state $|\psi^\BH_c\rangle$ is in fact close to the product state
\begin{equation}\label{eq:psi_c}
|\psi_c\rangle=\prod_{j=1}^{L/2}\hat{a}^{\dagger}_{\uparrow,j}\hat{a}^{\dagger}_{\downarrow,j+L/2}|0\rangle
 = \left|\uparrow\uparrow\dots\uparrow\downarrow\downarrow\dots\downarrow \right\rangle
\end{equation}
where $\left|0\right\rangle$ is the vacuum state.  The larger the interaction strengths in $\hat{H}_{\text{BH}}$, that is, the deeper the system is in the Mott-insulator phase, the greater the overlap of $\left|\psi^\BH_c\right\rangle$ and $\left|\psi_c\right\rangle$.

In principle, one obtains the XXZ model or the $\mathfrak{t}$-$J$ model within second-order perturbation theory as the effective model for the BH model for the sector of single-site \emph{bosonic} states $\{\left|\uparrow\right\rangle,\left|\downarrow\right\rangle\}$ or $\{\left|0\right\rangle,\left|\uparrow\right\rangle,\left|\downarrow\right\rangle\}$, respectively. Formally one gets from the original to the effective model via a unitary Schrieffer-Wolff transformation $e^{i\hat{\mathcal{S}}}$ followed by a projection to the aforementioned subspace. So the correspondence between the \emph{spin} $\left|\sigma\right\rangle$ and \emph{bosonic atom} $\left|\sigma\right\rangle$ is not $1:1$ -- one has corresponding perturbative corrections on top due to the unitary transformation \cite{Barthel2009}.  Thus, if one wants to study the analog of the XXZ domain-wall dynamics in the BH model, one should not start from the state $\left|\psi_c\right\rangle$, but take the perturbative corrections into account. If one did not, one would have nontrivial dynamics also in the ``fully polarized'' regions that are not influenced by the domain wall. In a bosonic state $\left|\uparrow\uparrow\uparrow\dots\uparrow\right\rangle$, for example, the BH dynamics is not trivial: Due to the hopping, states with $n_{\uparrow,j}\neq 1$ get populated (also, the boundary acts as a distortion). This also leads to entanglement growth in this supposedly trivial state. One can take into account the perturbative corrections very easily, by choosing the initial state, as described above, to be the ground state $|\psi^\BH_c\rangle$ of the BH model with a strong chemical potential for `$\uparrow$' bosons on the left half and for `$\downarrow$' bosons on the right half. This state is the actual counterpart of the spin domain-wall state $|\psi_c\rangle=\left|\uparrow\uparrow\dots\uparrow\downarrow\downarrow\dots\downarrow\right\rangle$ in the XXZ chain and the dynamics far away from the center is trivial then as it should be. With the Schrieffer-Wolff transformation $e^{i\hat{\mathcal{S}}}$, the correspondence between the states is
\begin{equation}\label{eq:SW_state}
 |\psi^\BH_c\rangle \approx e^{-i\hat{\mathcal{S}}}|\psi_c\rangle.
\end{equation}

The dominant defects occurring in the different experimental preparation schemes, as described in the introduction, are holes
\begin{equation}\label{eq:stateHoleBH}
|\psi^\BH_{h}\rangle = \hat{b}_{\uparrow,j_h}\left|\psi^\BH_c\right\rangle
\end{equation}
and spin flips
\begin{equation}\label{eq:stateFlipBH}
|\psi^\BH_{f}\rangle = \hat{b}^{\dagger}_{\downarrow,j_{f}}\hat{b}_{\uparrow,j_{f}}\left|\psi^\BH_c\right\rangle,
\end{equation}
where, in this paper, the defects are initially located in the left half of the system ($1<j_h,j_{f}<L/2$) without loss of generality. These types of defects naturally arise in the initial-state preparation or can simply be prepared deterministically in order to investigate their effects.

\begin{figure}[]
\includegraphics[width=0.98\columnwidth]{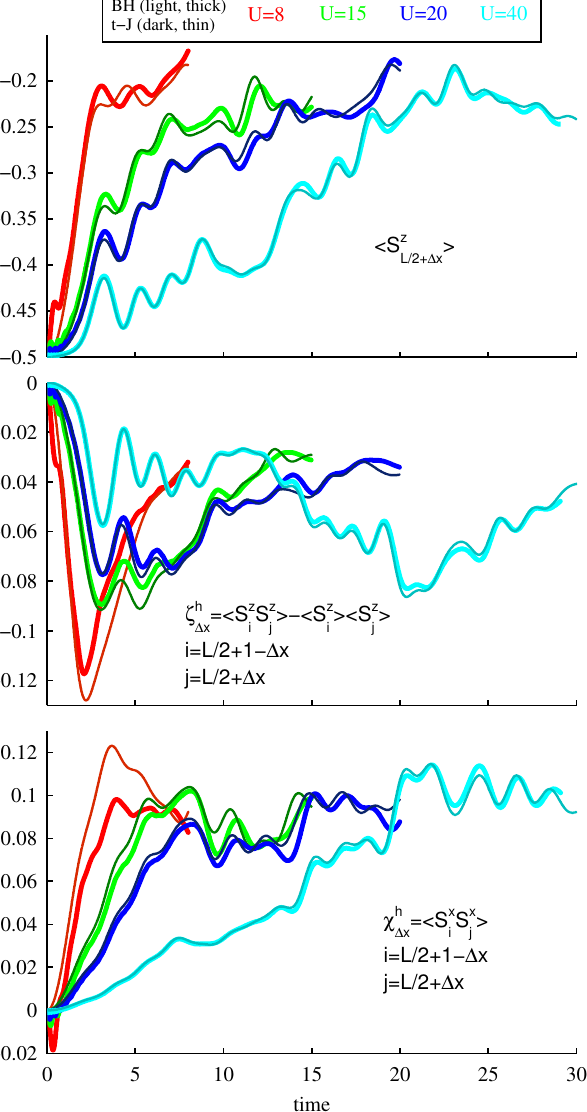}
\caption{(Color online)
Comparison of the dynamics in the full BH model \eqref{eq:BH} [light, thick lines] and the corresponding $\mathfrak{t}$-$J$ model \eqref{eq:tJ} [dark, thin lines] for an initial state with a hole at $j_h=L/2-4$, where $L=20$ and $\mathfrak{t}=1$. Here, $\Delta x =1$ for all observables.  Even for the smallest $U$ ($=8$), there is good agreement between the results of the BH and $\mathfrak{t}$-$J$ models. With increasing $U$, the agreement improves and the time from which on deviations become appreciable increases.
}
\label{fig:Fig13}
\end{figure}
\section{Convergence of BH dynamics to the \texorpdfstring{$\mathfrak{t}$-$J$}{t-J} dynamics and DMRG specifics}
For the reasons given in Sect.~\ref{sec:XXZmodel}, in the following, the analysis will be restricted to the isotropic case, where $\mathfrak{t}_\uparrow=\mathfrak{t}_\downarrow\equiv\mathfrak{t}$ and $U_{\uparrow}=U_{\downarrow}=V\equiv U$. The resulting isotropic two-species BH model \eqref{eq:BH} is found to map faithfully onto the $\mathfrak{t}$-$J$ model \eqref{eq:tJ} for $U\gtrsim 8$ ($\mathfrak{t}=1$).
Fig.~\ref{fig:Fig13} shows a comparison of the BH- and $\mathfrak{t}$-$J$-model results for the observable $\langle\hat{S}^z_{L/2+\Delta x}\rangle$ as well as the connected two-point correlation functions $\langle\hat{S}^z_{i}\hat{S}^z_{j}\rangle-\langle\hat{S}^z_{i}\rangle\langle\hat{S}^z_{j}\rangle$ and $\langle\hat{S}^x_{i}\hat{S}^x_{j}\rangle$ (note that $\langle\hat{S}^x_{i}\rangle=0$ for all times) for an initial hole state where the hole is located at $L/2-4$ and $\Delta x=1$. For all considered observables, the agreement is good for all values of $U\gtrsim 8$, and matches remarkably well for larger $U$, as is expected.

Although it is not an exact correspondence, we used here for the Bose-Hubbard model
\begin{equation}\label{eq:BH_approxSpinOp}\textstyle
	\frac{1}{2}\sum_{\sigma\sigma'}\hat{b}^\dag_{\sigma,j}[\hat{\sigma}^\alpha]_{\sigma\sigma'}\hat{b}^{\phantom{\dag}}_{\sigma',j} 
\end{equation}
as the counterpart of the observable $\hat{S}^\alpha_j$ in the $\mathfrak{t}$-$J$ model. In analogy to the relation \eqref{eq:SW_state} of states in the original (BH) and the effective model ($\mathfrak{t}$-$J$) which is ``simulated'' in the original model, the correct counterpart of an observable $\hat{O}$ of the effective model is $e^{-i\hat{\mathcal{S}}}\hat{O}e^{i\hat{\mathcal{S}}}$ for the original model.
Simply employing $\hat{O}$ also in the original model, as we did in this case by using the expression \eqref{eq:BH_approxSpinOp} instead of $e^{-i\hat{\mathcal{S}}}\hat{S}^\alpha_i e^{i\hat{\mathcal{S}}}$ causes an error $\mathcal O( \hat{\mathcal{S}})=\mathcal O(\mathfrak{t}/U)$ in the expectation values.
A thorough discussion concerning these issues can be found in Ref.~\cite{Barthel2009}.

All simulations are carried out using tDMRG \cite{Vidal2004,White2004,Daley2004} in the Krylov approach \cite{Feiguin2005,Garcia-Ripoll2006-8,McCulloch2007-10} (see also \cite{Schmitteckert2004-70}), where we compute each Krylov vector as a separate matrix product state. In the DMRG framework, one can control the accuracy of the simulation using the so-called \emph{truncation} or \emph{fidelity threshold} \cite{White1992,White1993,Schollwoeck2005} which puts an upper bound on the norm-distance between the exactly evolved state and the approximately evolved state of the simulation. For the results of the main text, we used a threshold of $10^{-6}$ for each time step and time steps of size $0.01/\mathfrak{t}$ and $0.1/\mathfrak{t}$ for the BH and the $\mathfrak{t}$-$J$ model, respectively. The convergence of the numerical results with respect to the fidelity threshold is demonstrated in the appendix.

\section{Domain-wall melting with and without defects}
As was described and numerically checked above, the two-species BH model $\hat{H}_{\text{BH}}$ maps onto the spin-$1/2$ XXZ model $\hat{H}_{\text{XXZ}}$ \cite{Kuklov2003,Duan2003,Ripoll2003,Altman2003,Barthel2009} or on the hard-core boson $\mathfrak{t}$-$J$ model \eqref{eq:tJ} for sufficiently large $U/\mathfrak{t}\gtrsim 8$. As this is the regime of experimental interest we can hence base the further analysis on simulations of the $\mathfrak{t}$-$J$ model.
The $\mathfrak{t}$-$J$-model parameters are set to $\mathfrak{t}=1$ and $U=15$ for a lattice of $L=320$ sites. The domain wall is located between sites $L/2$ and $L/2+1$ and, in the following, the cases of the clean domain-wall state $|\psi_c\rangle$ [Eq.~\eqref{eq:psi_c}], a domain wall with a hole defect at site $j_h=L/2-16$, $\hat{a}_{\uparrow,j_h}|\psi_c\rangle$, and a domain wall with a spin-flip defect at site $j_{f}=L/2-8$, $\hat{S}^-_{j_f}|\psi_c\rangle$, are investigated beginning with the magnetization profiles as shown in Fig.~\ref{fig:Fig3}.

\subsection{Magnetization profiles}
\begin{figure}[]
\includegraphics[width=\columnwidth]{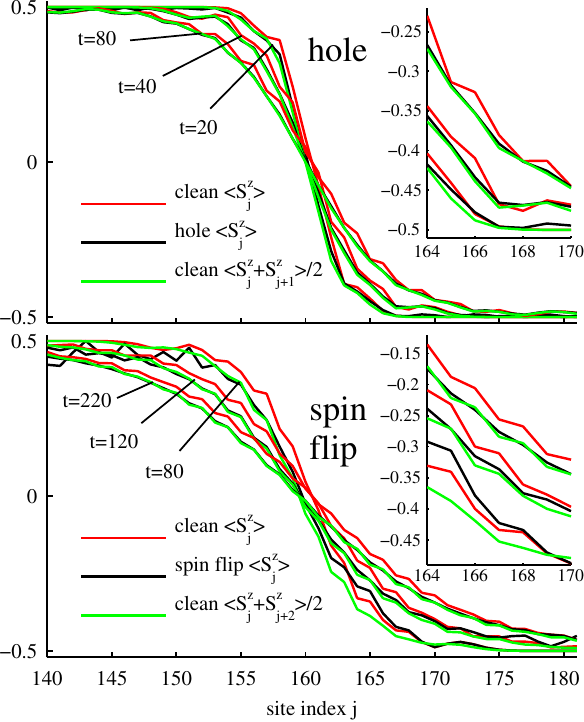}
\caption{(Color online)
Investigation of the influence of hole and spin-flip defects on the magnetization profile ($\mathfrak{t}=1$, $U=15$, $L=320$, $j_h=L/2-16$, $j_f=L/2-8$).  It is observed that at long times, the magnetization profile in the hole case matches its corresponding average quantity \eqref{eq:superpose} due to the superposition hypothesis (see text) very well, whereas the correspondence is not as good in the spin-flip case. This is due to the fact that the hole completely passes through the domain wall while the spin flip does not.
}
\label{fig:Fig5}
\end{figure}
\begin{figure*}[]
\includegraphics[width=0.88\textwidth]{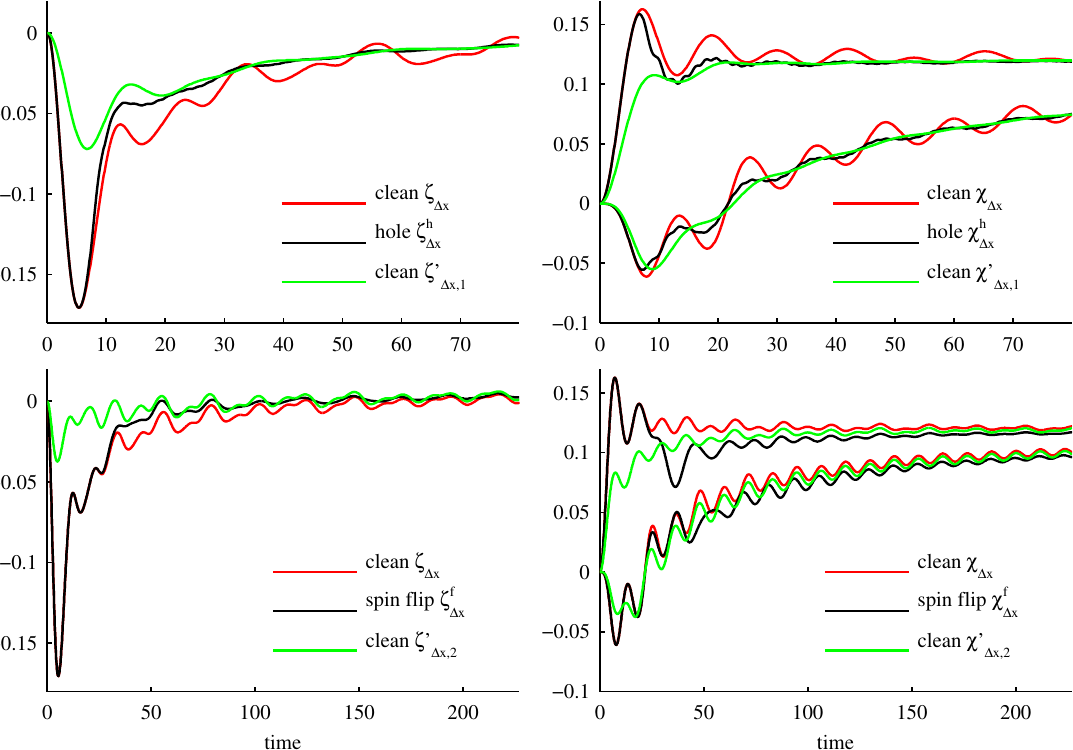}
\caption{(Color online)\label{fig:Fig67}
Left: Connected $S^z$--$S^z$ correlators \eqref{eq:correl} with $\Delta x=1$ ($\mathfrak{t}=1$, $U=15$, $L=320$, $j_h=L/2-16$, $j_f=L/2-8$).  The results show that the correlator $\zeta^h_{\Delta x}$ for the hole case deviates quite strongly from the same correlator $\zeta_{\Delta x}$ in the clean case. But, after the point in time where the hole has passed the domain wall, $\zeta^h_{\Delta x}$ agrees very well with $\zeta'_{\Delta x,1}$ [Eq.~\eqref{eq:zetaP}] which is computed from the clean correlator $\zeta^h_{\Delta x}$ by superimposing results for a small spatial shift. The coincidence of $\zeta^f_{\Delta x}$ and $\zeta'_{\Delta x,2}$ is not as good, indicating that holes have much less influence on the spin dynamics than spin-flip defects. 
Right: The same conclusion holds for the $S^x$--$S^x$ correlators.  Also here, $\chi^h_{\Delta x}$ agrees very well with $\chi'_{\Delta x,1}$ [Eq.~\eqref{eq:chiP}], while deviations of $\chi^f_{\Delta x}$ from $\chi'_{\Delta x,2}$ are still appreciable and comparable to the deviation from $\chi_{\Delta x}$.}
\end{figure*}
The clean case is not surprising and exhibits the known domain-wall melting dynamics associated with the Heisenberg XXZ model \cite{Gochev1977-26,Gochev1983-58,Antal1999,Hunyadi2004,Gobert2005,Mossel2010-12,Jesenko2011-84,Zauner2012}. A noteworthy feature of this domain-wall melting process in the clean case is a nearest-neighbor \emph{beating} mechanism (synchronized opposing oscillations of the magnetizations on neighboring sites) that persists even at later times and that one can make out in Fig.~\ref{fig:Fig3} and clearly see in Fig.~\ref{fig:Fig4}. Corresponding animations are available online \cite{epaps}. This feature, which causes short magnetization steps, is noticeably missing in the hole case, where the beating is strongly reduced and the short steps in the magnetization are smoothened out. At long times a further notable feature consists in (short) magnetization plateaus around the fronts of the domain wall \cite{Hunyadi2004,Zauner2012} that will also be smoothened in the presence of hole defects.
The spin-flip defect does not smoothen out the beating but it does have an effect on it nevertheless, as also shown in Fig.~\ref{fig:Fig4}. Figure~{\ref{fig:Fig3} also shows the significant difference in the velocities of the hole ($2\mathfrak{t}=2$) and that of the spin flip ($4\mathfrak{t}^2/U=4/15$) which is a manifestation of the spin-charge separation \cite{Luttinger1963,Haldane1981,Giamarchi}. In Fig.~\ref{fig:Fig3} it is difficult to pinpoint the influence of the defects on the domain wall, but Fig.~\ref{fig:Fig4} indicates for example that there is a resulting spatial shift of the magnetization profile. In Fig.~\ref{fig:Fig4} this shift is the reason why magnetization profiles of the evolved defect states at different times do not intersect the $\langle\hat{S}^z_j\rangle=0$ line at the same point. For sufficiently large times, the magnetization profile is, in comparison to the clean domain-wall evolution, shifted by about 0.5 sites in the hole-case and by 1 site in the spin-flip case.

\subsection{Explanation for the smoothening effect and spatial shifts}
The observed spatial shifts and smoothening effects can be understood as follows.
If one looks at the system at some long time $t$, at which the right-traveling part of the defect is assumed to have passed thorough the domain wall, one can express the time-evolved wave function $|\psi\rangle$ as a superposition of two approximately orthogonal states:
\begin{equation}\label{eq:superposition}
|\psi\rangle=\frac{1}{\sqrt{2}}(|\psi_\ell\rangle+|\psi_r\rangle).
\end{equation}
Here, $|\psi_\ell(t)\rangle$ describes a state with a defect traveling to the left, and thus the defect never interacts with the domain wall, and $|\psi_r(t)\rangle$ describes a state with a defect traveling to the right and that has already interacted with and passed through the wall.  Now in the case of the hole defect, due to the absence of one `$\uparrow$' boson, it is expected that the domain wall in $|\psi_r(t)\rangle$ is shifted by a single site towards the left, while in the case of the spin flip, not only is one `$\uparrow$' boson missing, but in its place we have an extra `$\downarrow$' boson, and thus the shift is expected to be by two sites to the left. The states $|\psi_{\ell,r}(t)\rangle$ that contain defect wave packets traveling to the left or right, respectively, are for sufficiently long times (approximately) orthogonal. This is due to the conservation of the particle number (hole case) or magnetization (spin-flip case) in the spatial region where the left-moving wave packet is supported. With this, one obtains at some site $j$ not too far away from the domain-wall region
\begin{align}\label{eq:superpose} \nonumber
\langle\psi|\hat{S}^z_j|\psi\rangle&=\frac{1}{2}(\langle\psi_\ell|\hat{S}^z_j|\psi_\ell\rangle+\langle\psi_r|\hat{S}^z_j|\psi_r\rangle) \\ 
&\approx \frac{1}{2}(\langle\psi_c|\hat{S}^z_j|\psi_c\rangle+\langle\psi_c|\hat{S}^z_{j+d}|\psi_c\rangle),
\end{align}
where $d=1$ ($2$) in the case where the defect is a hole (spin flip) and $|\psi_c(t)\rangle$ is the evolved wave function for the clean domain-wall state. Figure~\ref{fig:Fig5} shows the magnetization profile for each of the hole and spin-flip states at three different points in time as compared to the corresponding magnetization profiles for the clean state $|\psi_c\rangle$ and the corresponding magnetization profiles due to the superposition as quantified in Eq.~\eqref{eq:superpose}.  In the case of the hole, there is great agreement between the magnetization profile of the hole state and Eq.~\eqref{eq:superpose}, especially for long times at which the hole has already passed through the domain wall (Fig.~\ref{fig:Fig3}).  Moreover, this averaging of two density profiles shifted by one site from each other \eqref{eq:superpose} explains well why the beating observed for the clean case in Fig.~\ref{fig:Fig4} is smoothened out in the case of the hole state: The beating consists of a synchronized opposing oscillation of the magnetizations on neighboring sites. Summing the magnetization profile and its one-site translate \eqref{eq:superpose}, the opposing oscillations basically cancel out. 
The remaining smaller deviations are beyond the simple ``classical'' shifting effect. They are due to the modification of the domain-wall dynamics caused by the passing hole. At the location of the passing hole, the spin-spin interaction is practically switched off for a short period of time. This alteration of the domain-wall evolution will reduce with increasing $U$, as the hole will then pass faster and faster through the domain wall (when viewed in time units of $1/J$).
In the case of the spin-flip defect, the superposition picture \eqref{eq:superposition} is still useful but not as powerful for explaining the deviations to the clean case. One observes that, even at longer times, the magnetization profile for the spin-flip state does not fully converge to the corresponding averaged profile of Eq.~\eqref{eq:superpose}. This is due to the fact that the spin-flip defect dynamics occurs on the same time scale as the domain-wall dynamics and that, at least up to the maximum simulated times, the spin flip does not completely pass through the domain wall. 
Besides this, it is clear that the beating is not reduced by the spin-flip defect because, according to the superposition hypothesis, one has to add magnetizations of next-nearest neighbor sites ($d=2$ in Eq.~\eqref{eq:superpose}) for which the beating oscillations are in sync.

\subsection{Correlation functions}
The above intuitive notion of a superposition of left- and right-moving defects works well when it comes to the magnetization profile. Additionally, one can see how it fares when considering experimentally relevant connected two-point correlators around the domain wall
\begin{subequations}\label{eq:correl}
\begin{align}
\zeta_{\Delta x}&:=\langle\hat{S}^z_{i}\hat{S}^z_{j}\rangle-\langle\hat{S}^z_{i}\rangle\langle\hat{S}^z_{j}\rangle
\quad\text{and}\\
\chi_{\Delta x}&:=\langle\hat{S}^x_{i}\hat{S}^x_{j}\rangle,
\end{align}
\end{subequations}
where $i=L/2+1-\Delta x$ and $j=L/2+\Delta x$. For clarity, $\zeta_{\Delta x}$ and $\chi_{\Delta x}$ will refer to the clean case, while in the case of a hole or a spin flip, both two-point correlators will be augmented with the superscript ``h'' or ``f'', respectively. Moreover, it is to be noted that, in this model, one always has $\langle\hat{S}^x_j\rangle=\langle\hat{S}^y_j\rangle=0$, hence the apparent difference in the definitions of $\zeta_{\Delta x}$ and $\chi_{\Delta x}$. Based on the superposition in Eq.~\eqref{eq:superposition}, the two-point correlators for the defect case should agree with
\begin{align}\nonumber
\zeta'_{\Delta x,d}:=&\frac{1}{2}\left(\langle\psi_c|\hat{S}^z_i\hat{S}^z_j|\psi_c\rangle+\langle\psi_c|\hat{S}^z_{i+d}\hat{S}^z_{j+d}|\psi_c\rangle\right) \\ \nonumber
&+\frac{1}{4}\left(\langle\psi_c|\hat{S}^z_i|\psi_c\rangle+\langle\psi_c|\hat{S}^z_{i+d}|\psi_c\rangle\right) \\ 
&\quad\,\times\left(\langle\psi_c|\hat{S}^z_j|\psi_c\rangle+\langle\psi_c|\hat{S}^z_{j+d}|\psi_c\rangle\right) \label{eq:zetaP}
\end{align}
and
\begin{align} 
\chi'_{\Delta x,d}&:=\frac{1}{2}\left(\langle\psi_c|\hat{S}^x_i\hat{S}^x_j|\psi_c\rangle+\langle\psi_c|\hat{S}^x_{i+d}\hat{S}^x_{j+d}|\psi_c\rangle\right), \label{eq:chiP}
\end{align}
respectively. As above, we have again $d=1$ for the hole case and  $d=2$ for the spin-flip case. As shown in Fig.~\ref{fig:Fig67}, $\zeta^h_{\Delta x}$ ($\chi^h_{\Delta x}$) agrees well with $\zeta'_{\Delta x,1}$ ($\chi'_{\Delta x,1}$) for longer times, and this behavior supports the idea that the hole indeed passes through the domain wall completely, leading to a smoothening effect as dictated by the superposition concept of Eq.~\eqref{eq:superposition}.  However, in the case of the spin flip, the explanatory power of this concept is again not as impressive.

\subsection{Particle density is independent of spin dynamics}
\begin{figure}[]
\includegraphics[width=0.97\columnwidth]{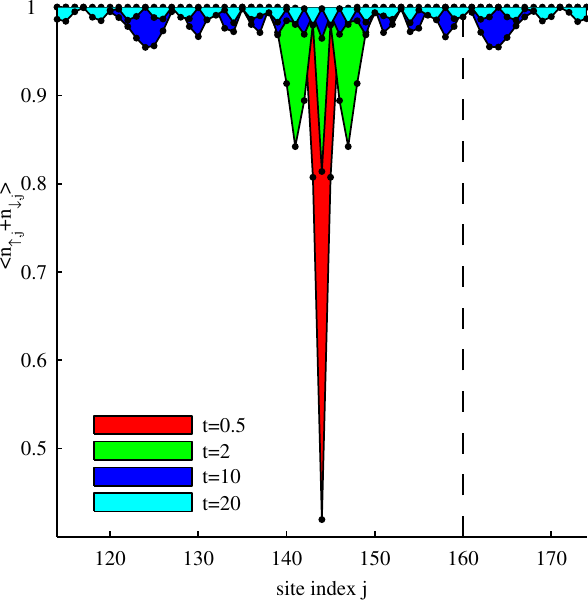}
\caption{(Color online)
Dynamics of the particle density $\langle \hat{n}_{\uparrow,j}+\hat{n}_{\downarrow,j}\rangle$ for the domain-wall state with a hole defect ($\mathfrak{t}=1$, $U=15$, $L=320$, $j_h=L/2-16$). The particle density is symmetric with respect to the initial position of the hole and shows now particular features at the domain wall (dashed line). As discussed in the text, the density dynamics is in fact completely independent of the spin dynamics (the converse is of course not the case).
}
\label{fig:Fig8}
\end{figure}
\begin{figure*}[p]
\includegraphics[width=\textwidth]{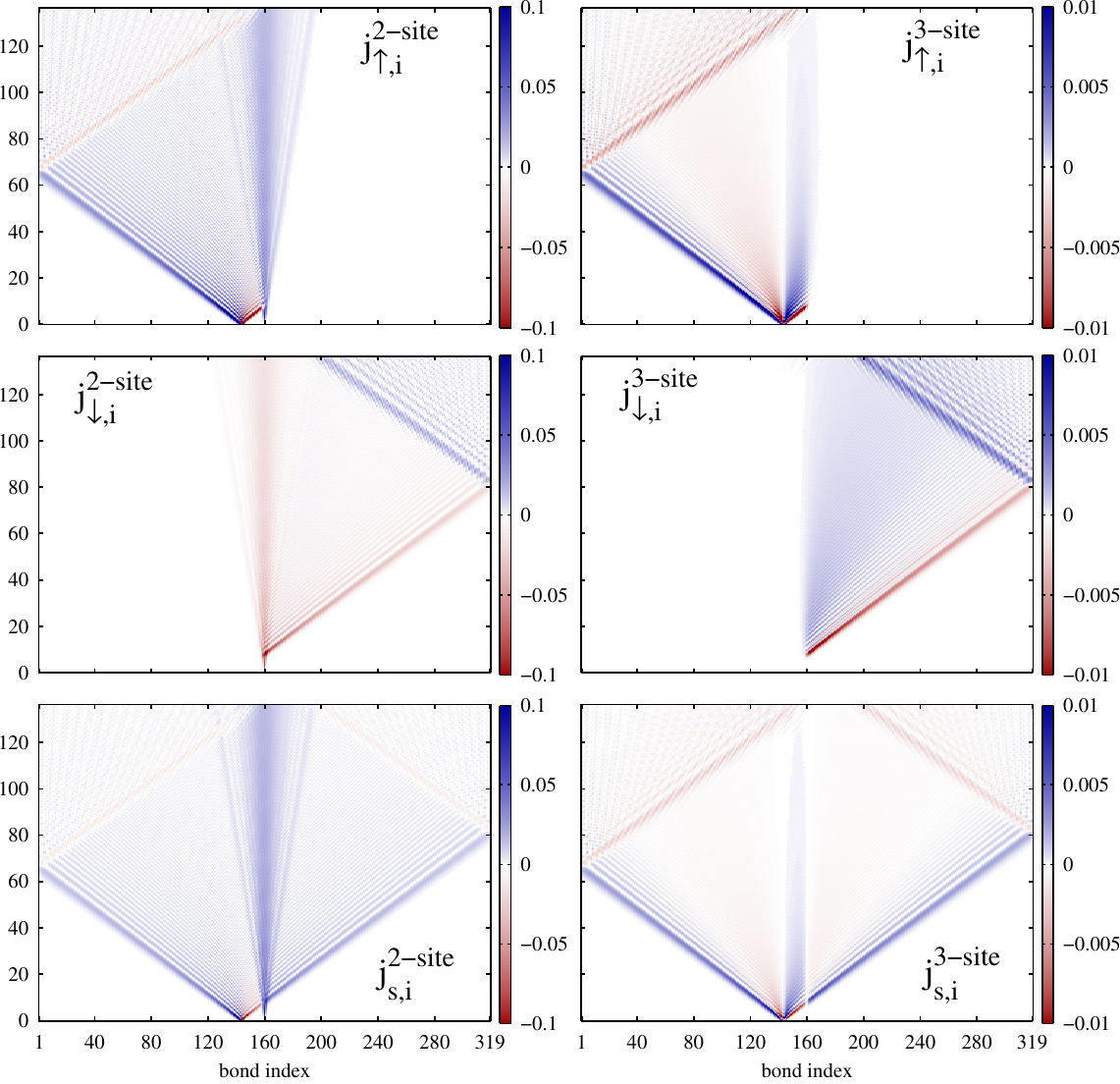}
\caption{(Color online)
Currents \eqref{eq:current} for a domain wall with a hole defect ($\mathfrak{t}=1$, $U=15$, $L=320$, $j_h=L/2-16$). The figures show from top to bottom the density currents for `$\uparrow$' bosons, `$\downarrow$' bosons, and the spin current. On the left, the contributions of the dominating two-site terms \eqref{eq:current2} are shown. The contributions \eqref{eq:current3} of the three-site hopping terms ($\hat{H}_{3-\text{site}}$ in the $\mathfrak{t}$-$J$ Hamiltonian \eqref{eq:tJ}) are given on the right. For the given $U$, they are suppressed by one order of magnitude. The suppression is stronger for larger $U$. 
}
\label{fig:Fig10}
\end{figure*}
Next, the particle density is considered. In the case of the clean domain-wall state and the case of a domain wall with a spin-flip defect, the particle density is simply constant with exactly one particle per site for all times. For the hole defect, one might naively expect some nontrivial effects, like reflection of the hole from the domain wall etc. However, as the Hamiltonian terms that change the particle density distribution are species-independent ($\mathfrak{t}_\uparrow=\mathfrak{t}_\downarrow$), the hole dynamics is completely independent of the spin dynamics. This is visualized in Fig.~\ref{fig:Fig8} where the initial position of the domain wall is marked by a dashed line. Indeed, $\langle\hat{n}_{\uparrow,j}+\hat{n}_{\downarrow,j}\rangle$ is symmetric around the initial position of the hole for all times and shows no special features in the domain-wall region.

\subsection{Quantification of higher-order effects by spin and density currents}
Finally, let us consider the spin and density (``charge'') currents during the dynamics. They correspond to specific short-range correlators which are in principle accessible in experiments. Besides offering another perspective on the evolution of the domain wall and the defects, we can use it to quantify the effect of the higher-order (three-site) hopping terms $\hat{H}_{3-\text{site}}$ in the effective model \eqref{eq:tJ}.  The density current $\hat{j}_{\sigma,i}$ for boson species `$\sigma$' at a bond ($i$,$i+1$), denoted by the bond index $i$, is defined as the time derivative of the total particle number $\sum_{j>i}\hat{n}_{\sigma,j}$ to the right of that bond.  For the $\mathfrak{t}$-$J$ model \eqref{eq:tJ}, one obtains:
\begin{align} \nonumber
\hat{j}_{\sigma,i}=& -\text{i}\mathfrak{t}_{\sigma}(\hat{b}^{\dagger}_{\sigma,i}\hat{b}_{\sigma,i+1}-h.c.) 
+\text{i}\frac{J_{\perp}}{2}(\hat{S}^{\sigma}_i\hat{S}^{-\sigma}_{i+1}-h.c.)  \\ \label{eq:current}
&+\hat{j}^a_{\sigma,i}+\hat{j}^b_{\sigma,i}+\hat{j}^c_{\sigma,i},
\end{align}
where
\begin{align*}
\hat{j}^a_{\sigma,i}=&-\frac{\text{i}\mathfrak{t}_{\sigma}^2}{V}(\hat{b}^{\dagger}_{\sigma,i-1}\hat{n}_{-\sigma,i}\hat{b}_{\sigma,i+1}
+\hat{b}^{\dagger}_{\sigma,i}\hat{n}_{-\sigma,i+1}\hat{b}_{\sigma,i+2}-h.c.),\\
\hat{j}^b_{\sigma,i}=&-\frac{\text{i}\mathfrak{t}_{\uparrow}\mathfrak{t}_{\downarrow}}{V}(\hat{b}^{\dagger}_{-\sigma,i-1}\hat{S}^{\sigma}_{i}\hat{b}_{\sigma,i+1}
+\hat{b}^{\dagger}_{\sigma,i}\hat{S}^{-\sigma}_{i+1}\hat{b}_{-\sigma,i+2}-h.c.),\\
\hat{j}^c_{\sigma,i}=&-\frac{2\text{i}\mathfrak{t}_{\sigma}^2}{U_{\sigma}}(\hat{b}^{\dagger}_{\sigma,i-1}\hat{n}_{\sigma,i}\hat{b}_{\sigma,i+1} 
+\hat{b}^{\dagger}_{\sigma,i}\hat{n}_{\sigma,i+1}\hat{b}_{\sigma,i+2}-h.c.).
\end{align*}
The spin current is then simply
\begin{equation}
\hat{j}_{s,i}=\frac{1}{2}\left(\hat{j}_{\uparrow,i}-\hat{j}_{\downarrow,i}\right).
\end{equation}
In Fig.~\ref{fig:Fig10} the two-site and three-site contributions to the charge and spin currents
\begin{align} \label{eq:current2}
\hat{j}^{2-\text{site}}_{\sigma,i}&=-\text{i}\mathfrak{t}_{\sigma}(\hat{b}^{\dagger}_{\sigma,i}\hat{b}_{\sigma,i+1}-h.c.)
+\text{i}\frac{J_{\perp}}{2}(\hat{S}^{\sigma}_i\hat{S}^{-\sigma}_{i+1}-h.c.) \\ \label{eq:current3}
\hat{j}^{3-\text{site}}_{\sigma,i}&=\hat{j}^a_{\sigma,i}+\hat{j}^b_{\sigma,i}+\hat{j}^c_{\sigma,i}
\end{align}
and 
\begin{equation}\label{eq:currentS}
\hat{j}^{m-\text{site}}_{s,i}=\hat{j}^{m-\text{site}}_{\uparrow,i}-\hat{j}^{m-\text{site}}_{\downarrow,i}
\end{equation}
with $m=2$ or $3$ are shown for $\mathfrak{t}=1$ and $U=15$. The currents offer another deeper look at the dynamics, visualizing the flow of particles  and magnetizations.  For the given parameters, the contributions of the effective three-site hopping terms is one order of magnitude below that of the two-site terms. Their effect decreases further for larger $U$.

\subsection{Multiple defects}
\begin{figure}[]
\includegraphics[width=\columnwidth]{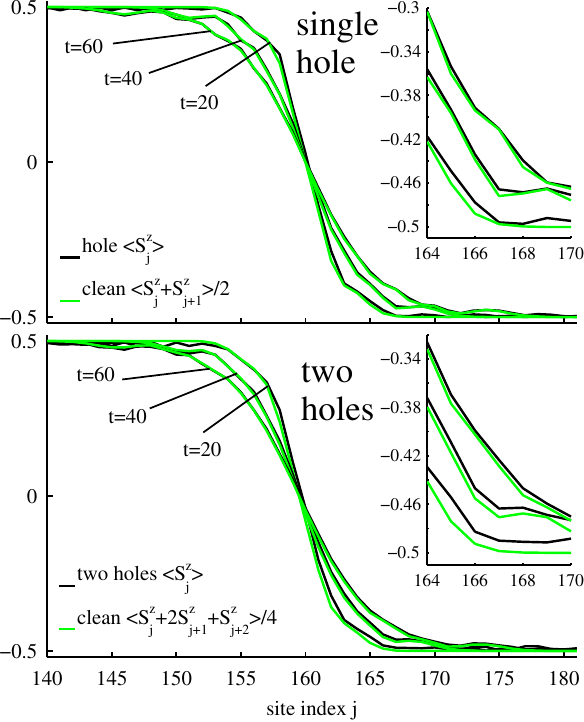}
\caption{(Color online)
Investigating the effect of two holes with initial positions $j_{h_1}=L/2-16$ and $j_{h_2}=L/2-8$ ($\mathfrak{t}=1$, $U=15$).  The magnetization profiles at long times indicate that, like in the single-hole case [\eqref{eq:superposition}], the two-hole dynamics can be approximated as a superposition of orthogonal states which correspond to spatial shifts of the clean domain-wall state; see the text and Eq.~\eqref{eq:twoholesuperposition}.
}
\label{fig:Fig14}
\end{figure}
Now that the effect of a single hole defect on the domain-wall evolution is understood, one may be interested in investigating, on the one hand, the effect of two simultaneously present holes, and on the other hand, whether or not such two hole defects interact with each other.  For times when the left- and right-moving parts of the holes are sufficiently separated, one can once again intuitively describe the system by a superposition of orthogonal states
\begin{equation}\label{eq:twoholesuperposition}
|\psi\rangle = \frac{1}{2}(|\psi_{\ell\ell}\rangle+|\psi_{\ell r}\rangle+|\psi_{r\ell}\rangle+|\psi_{rr}\rangle),
\end{equation}
where $|\psi_{\ell\ell}\rangle$ describes two holes moving to the left, and thus they never interact with the domain wall, $|\psi_{\ell r}\rangle$ ($|\psi_{r\ell}\rangle$) is the state where the left (right) hole is moving to the left and never interacts with the wall while the right (left) hole has passed through the wall, shifting it by one site to the left, and $|\psi_{rr}\rangle$ describes the state where both holes have traveled to the right and passed through the domain wall, shifting it by two sites to the left.  This leads to the following magnetization profile for the two-hole state:
\begin{align}\label{eq:twoholesuperpose} \nonumber
\langle\psi|\hat{S}^z_j|\psi\rangle &=  \frac{1}{4}\left(\langle\psi_{\ell\ell}|\hat{S}^z_j|\psi_{\ell\ell}\rangle+\langle\psi_{\ell r}|\hat{S}^z_j|\psi_{\ell r}\rangle\right. \\ \nonumber
&\left.\quad\quad+\langle\psi_{r\ell}|\hat{S}^z_j|\psi_{r\ell}\rangle+\langle\psi_{rr}|\hat{S}^z_j|\psi_{rr}\rangle\right) \\ \nonumber
&\approx\frac{1}{4}\left(\langle\psi_c|\hat{S}^z_j|\psi_c\rangle+2\langle\psi_c|\hat{S}^z_{j+1}|\psi_c\rangle\right. \\ 
&\left.\quad\quad+\langle\psi_c|\hat{S}^z_{j+2}|\psi_c\rangle\right)
\end{align}
The magnetization profiles for a single-hole at site $j_h=L/2-16$ and two-holes at sites $j_{h_1}=L/2-16$ and $j_{h_2}=L/2-8$ are shown in Fig.~\ref{fig:Fig14} along with the corresponding curves due to Eqs.~\eqref{eq:superpose} and \eqref{eq:twoholesuperpose} at three different points in time. One observes that, especially at longer times, the magnetization profile of the two-hole state matches remarkably well the curve due to Eq.~\eqref{eq:twoholesuperpose}. The smaller deviations beyond this effect, are roughly proportional to the number of holes and decrease when $\mathfrak{t}/J$ is increased (larger $U$) as discussed in the following.

\subsection{Reducing the effects of holes by increasing \texorpdfstring{$\mathfrak{t}/J$}{t\textfractionsolidus J}}
\begin{figure}[]
\includegraphics[width=\columnwidth]{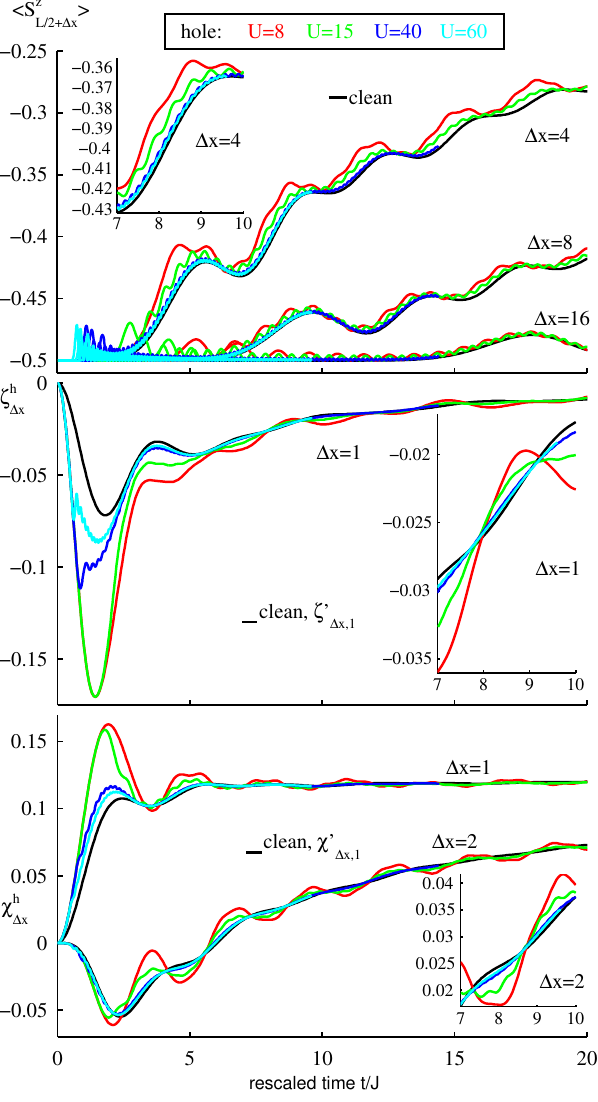}
\caption{(Color online)
Comparison of the effect of a hole defect on two-point correlation functions in the domain-wall dynamics ($\mathfrak{t}=1$, $L=320$, $j_h=L/2-16$) at different values of the on-site repulsion $U$ for the $\mathfrak{t}$-$J$ model \eqref{eq:tJ}. The greater the value of $U$, the faster the hole propagates relative to the domain-wall front, and thus the shorter the interaction time between the hole and the domain wall, which in turn leads to greater agreement between the hole-case dynamics and the approximation due to the superposition hypothesis \eqref{eq:twoholesuperposition}.
}
\label{fig:Fig9}
\end{figure}
The effect of holes depends on the relative velocity of the holes with respect to the domain wall. A relatively faster hole has a smaller effect on the domain-wall dynamics, as the interaction time between the hole and the domain wall is smaller in such a case.  Alternatively, a relatively slower hole will have more time to distort the dynamics of the domain wall, and thus the dynamics will deviate stronger from the superposition behavior such as that described by Eq.~\eqref{eq:superpose}. Figure~\ref{fig:Fig9} shows the magnetization profiles and two-point correlators over time for the single-hole state ($j_h=L/2-16$) for different values of $U$.  One sees that the smaller $U$ is, and thus the slower the hole is relative to the domain-wall melting, the larger the deviation of the above observables from their superposition curves given in Eq.~\eqref{eq:superpose} for the magnetization profile and Eqs.~\eqref{eq:zetaP} and \eqref{eq:chiP} for the two-point correlators.  On the other hand, for very large $U=60$, the agreement between the magnetization profile and Eq.~\eqref{eq:superpose}, and between $\zeta^h_{\Delta x}$ ($\chi^h_{\Delta x}$) and $\zeta'_{\Delta x,1}$ ($\chi'_{\Delta x,1}$) is excellent after a certain short time $\sim1/\mathfrak{t}$ corresponding to the phase where the hole passes the domain-wall region.

\section{Conclusion}
The numerical simulations and the analysis of disturbances due to defects that we have provided in this paper give useful insights concerning future experiments using ultracold atomic gases to simulate the dynamics of quantum magnets.
Specifically, we have investigated domain-wall melting in the two-species Bose-Hubbard model in the presence of hole and spin-flip defects. For large on-site repulsions, the model maps to a hard-core boson $\mathfrak{t}$-$J$ model with particular effective three-site hopping terms. The study is based on tDMRG calculations using the Krylov approach. It is concluded, through measurements of magnetization profiles and two-point correlators, that a domain wall with a single hole defect evolves into a superposition of two (approximately) orthogonal states, where the domain-wall melting becomes equivalent to that of two domain walls, one of which is shifted towards the initial position of the hole by one site. The situation of multiple holes can be described in a similar manner. The leading effect of holes hence corresponds to a certain averaging of spatially shifted observables that can be taken account of. Further smaller deviations due to holes diminish with increasing repulsion $U$ as the hole dynamics gets faster and faster in comparison to the domain-wall evolution.
Whereas hole defects are in this sense not so problematic, the effect of spin-flip defects is more severe as they evolve on the same time scale as the domain wall itself. Although it is still useful, this limits the explanatory power of the superposition picture for spin-flip defects. For the experimental investigations this has implications on the preparation of the initial states. In particular, our results suggest that the second preparation scheme (see introduction), based on cooling with species- and position-dependent chemical potentials, should be favorable over the first scheme that is based on inducing spin-flips in parts of the system.

\section{Acknowledgments}
The authors are grateful to Immanuel Bloch, Marc Cheneau, Christian Gross, and Takeshi Fukuhara (all of LMU Physik and MPQ Garching) for fruitful and insightful discussions. This work has been supported through the FP7/Marie-Curie grant 321918 (J.C.H.), the Australian Research Council grant CE110001013 (I.P.M.), and DFG FOR 801 (U.S. and J.C.H.).

\appendix*

\begin{figure*}[p]
\begin{minipage}{\columnwidth}
\includegraphics[width=\columnwidth]{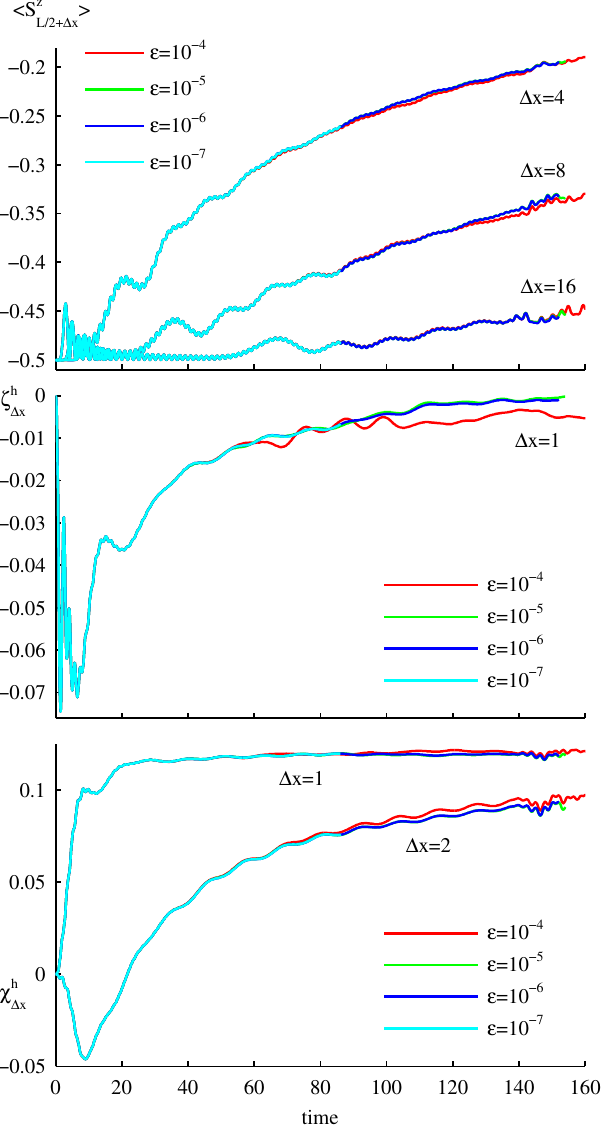}
\caption{\label{fig:Fig11}
 (Color online)
Convergence of tDMRG results with respect to the fidelity threshold $\epsilon$, for magnetizations and two-point correlators \eqref{eq:correl} for the $\mathfrak{t}$-$J$ model specified in Eq.~\eqref{eq:tJ} ($L=320$, $\mathfrak{t}=1$, $U=15$). Here, the initial state is the domain wall with a hole defect at site $L/2-1$. Good convergence is achieved at a fidelity threshold of $10^{-6}$.
}
\end{minipage}
\hspace{0.02\textwidth}
\begin{minipage}{\columnwidth}
\includegraphics[width=\columnwidth]{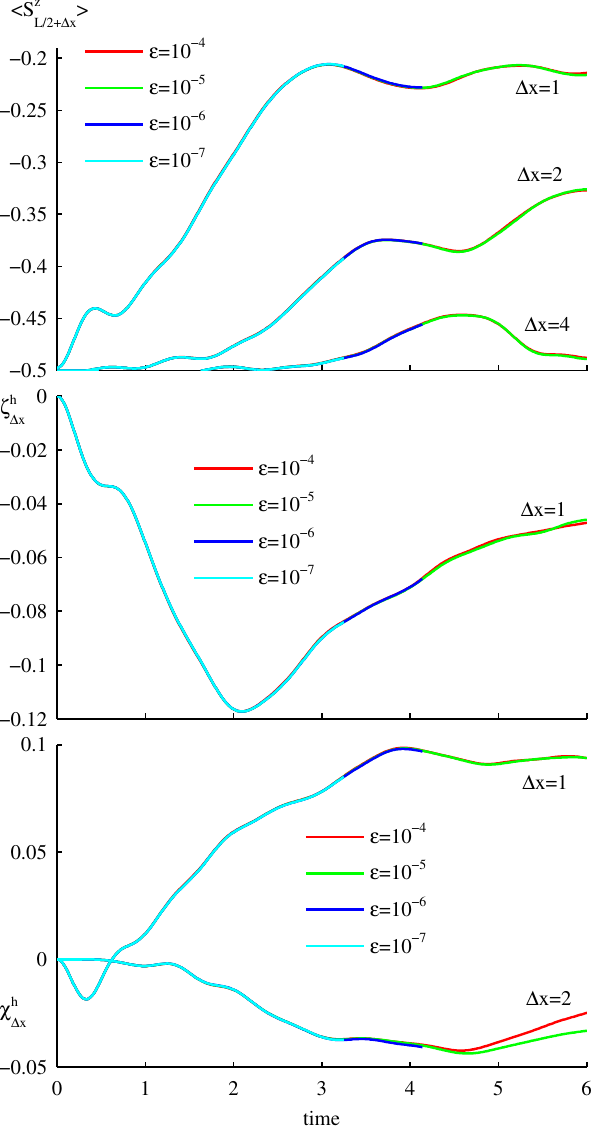}
\caption{\label{fig:Fig12}(Color online)
Convergence of tDMRG results with respect to the fidelity threshold $\epsilon$, for magnetizations and two-point correlators \eqref{eq:correl} for the two-species BH model specified in Eq.~\eqref{eq:BH} ($L=20$, $\mathfrak{t}=1$, $U=8$). The initial state is the domain wall with a hole defect at site $L/2-1$ [Eq.~\eqref{eq:stateHoleBH}]. Good convergence is achieved at a fidelity threshold of $10^{-6}$.
}
\end{minipage}
\end{figure*}

\section{Convergence of the DMRG simulations}
All simulations in this paper are carried out using tDMRG \cite{Vidal2004,White2004,Daley2004} in 
the Krylov approach \cite{Feiguin2005,Garcia-Ripoll2006-8,McCulloch2007-10} (see also \cite{Schmitteckert2004-70}) with time steps of a certain size $\Delta t$. In the tDMRG, the evolved many-body state is approximated as a so-called matrix product state at all times which is achieved by repeated truncations of small Schmidt coefficients. The accuracy of the simulation is controlled using a threshold $\epsilon$ on the fidelity loss due to truncations \cite{White1992,White1993,Schollwoeck2005}. Let $|\psi_t\rangle$ be the state for time $t$. In every time step, we apply the Hamiltonian $\hat{H}$ multiple times to $|\psi_t\rangle$, to obtain matrix product state representations of the Krylov vectors $\{|\psi_t\rangle,\hat{H}|\psi_t\rangle,\hat{H}^2|\psi_t\rangle,\dots\}$. Controlling errors due to DMRG truncations of the Krylov vectors and due to a restriction on the number of used Kryolv vectors, we implement the time evolution in the Krylov subspace to obtain a new matrix product state $|\psi_{t+\Delta t}\rangle$ such that $r^2:=\|\hat{U}_{\Delta t}|\psi_t\rangle-|\psi_{t+\Delta t}\rangle\|^2/\|\hat{U}_{\Delta t}|\psi_t\rangle+|\psi_{t+\Delta t}\rangle\|^2 < \epsilon$, where $\hat{U}_{\Delta t}$ is the (exact) time-evolution operator for a single time step. For the computation of a bound on $r$, we use some very conservative assumptions on the decay of the coefficients in the expansion of the evolved state in the Krylov basis.

The size of the time step was chosen such that the number of required Krylov vectors was roughly 10. In particular, we chose $\Delta t=0.01/\mathfrak{t}$ and $\Delta t=0.1/\mathfrak{t}$ for the Bose-Hubbard (BH) and the $\mathfrak{t}$-$J$ model, respectively. For all analyzed observables one should ensure convergence with respect to the fidelity threshold $\epsilon$. As described in the following we determined these parameters such that the data presented in the figures is quasi-exact. 

For the $\mathfrak{t}$-$J$ model, a lattice of $L=320$ sites was used and the results presented in the main text are based on a fidelity threshold of $\epsilon=10^{-6}$.  In order to check for convergence, several runs are carried out at different $\epsilon$ for the single-hole state where the hole is located at $L/2-1$ (this state is found to be the most challenging numerically among all initial states simulated) at $U=15$ and $\mathfrak{t}=1$.  Once again, the observable $\langle\hat{S}^z_{L/2+\Delta x}\rangle$ and the two-point correlators $\zeta^h_{\Delta x}$ and $\chi^h_{\Delta x}$ [Eq.~\eqref{eq:correl}] for various $\Delta x$ are taken into account, and as Fig.~\ref{fig:Fig11} shows, very good convergence is achieved at a fidelity threshold of $10^{-6}$.

Furthermore, in order to validate the comparison in Fig.~\ref{fig:Fig13}, one must ascertain the convergence of the corresponding BH-model results, where a fidelity threshold of $10^{-6}$ is also used.  Fig.~\ref{fig:Fig12} shows the observable $\langle\hat{S}^z_{L/2+\Delta x}\rangle$ and the two-point correlators $\zeta^h_{\Delta x}$ and $\chi^h_{\Delta x}$ at various $\Delta x$, and, indeed, a fidelity threshold of $10^{-6}$ exhibits very good convergence.

\bibliographystyle{prsty}

\end{document}